\documentclass[preprint]{aastex}
\usepackage{color}

\usepackage{amsmath}
\usepackage{natbib}
\usepackage{graphicx}
\usepackage{epstopdf}
\usepackage{color}
\usepackage{soul}

\slugcomment{Diffuse coronae}
\shorttitle{Properties of diffuse coronae}
\shortauthors{Soko{\l}owska et al.}
\begin{document}
\title{Diffuse coronae in cosmological simulations \\
    of Milky Way--sized galaxies}

\author{A. Soko{\l}owska\altaffilmark{1}, L. Mayer\altaffilmark{1}, A. Babul\altaffilmark{2}, P.Madau\altaffilmark{1,3}, \& S.Shen\altaffilmark{4}}

\email{alexs@physik.uzh.ch}

\altaffiltext{1}{Center for Theoretical Astrophysics and Cosmology, University of Zurich, Winterthurerstrasse 190, Zurich, Switzerland.}
\altaffiltext{2}{Department of Physics and Astronomy, University of Victoria, Elliot Bldg, 3800 Finnerty Rd, Victoria, Canada.}
\altaffiltext{3}{Department of Astronomy and Astrophysics, 1156 High Street, University of California, Santa Cruz CA 95064, USA.}
\altaffiltext{4}{Kavli Institute for Cosmology, University of Cambridge, Madingley Road, Cambridge CB3 0HA, UK.}

\begin{abstract}

We investigate the properties of halo gas using three cosmological `zoom-in' simulations of realistic Milky Way-galaxy analogs with  varying sub-grid physics. In all three cases, the mass of hot ($T > 10^6$ K) halo gas is $\sim 1\%$ of the host's virial mass. { The X-ray luminosity of two of the runs is consistent with observations of the Milky Way, while the third simulation is X-ray bright and resembles more closely a very massive, star-forming spiral.} Hot halos extend to 140 kpc from the galactic center and are surrounded by a bubble of warm--hot ($T = 10^5 - 10^6$K) gas that extends to the virial radius.  Simulated halos agree well outside 20--30 kpc with the $\beta$--model of \cite{Miller:2013aa} based on OVII absorption and OVIII emission measurements.  Warm--hot and hot gas contribute up to $80\%$ of the total gas reservoir, and contain nearly the same amount of baryons as the stellar component. The mass of warm--hot and  hot components falls into the range estimated for $L^*$ galaxies.   With key observational constraints on the density of the Milky Way corona being satisfied, we show that concealing of the ubiquitous warm--hot baryons, along with the ejection of just $20-30 \%$ of the diffuse gas out of the potential wells by supernova--driven outflows, can solve the "missing baryon problem". The recovered baryon fraction within 3 virial radii is $90\%$ of the universal value.  With a characteristic density of $\sim 10^{-4}$ cm$^{-3}$ at $50-80$ kpc, diffuse coronae meet  the requirement for fast and complete ram--pressure stripping of the gas reservoirs in dwarf galaxy satellites.

\end{abstract}


\section{Introduction}
\label{sec:intro}
Spiral galaxies are embedded in coronae of diffuse gas that connect intergalactic medium (IGM) 
to the star-forming disks of galaxies \citep{Putman:2012aa}. In the absence of any feedback processes, the corona 
is supplied with accreting gas from intergalactic medium and in turn, the cooling gas from corona provides fresh fuel for star formation. In the 
presence of feedback, however, outflows from the galaxy also contribute to the corona and may even expel the coronal gas beyond the galactic halo. Thus the ebb and tide 
of the coronal gas is part and parcel of the galaxy formation process. 

High--resolution cosmological simulations performed with hydrodynamical zoom--in codes have made significant progress in reproducing a wealth of galaxies' 
properties in various mass ranges (dwarfs: \cite{Shen:2014}; spiral galaxies: \cite{Guedes:2011aa}, \cite{Marinacci:2014}, \cite{ Agertz:2014}, \cite{Roskar:2014}; 
massive early--types: \cite{Argo:2014}), yet the interplay between cooling and heating, and the exchange of mass between galaxies and IGM remain open issues.  \cite{Crain:2013} inspected gas around massive disc galaxies reproducing X--ray scalings and found that electron densities of hot coronae are dominated by IGM accretion. Recently, the properties of gaseous galactic halos were characterized in massive $(10^{12}M_{\odot})$ halos at $z=2$ by \cite{Nelson:2015} who also pinpoint that inflow stream morphologies become continuously more complex with better numerical resolution.  In the Milky Way regime, \cite{Marinacci:2014G} emphasized the importance of galactic winds in shaping the structure of circumgalactic medium. These studies have only just scratched the surface in terms of understanding the nature and properties of galactic coronae in simulated galaxies. Such coronae can potentially yield a wealth of complementary constraints on the fidelity of the simulations, as well as
on the details of the sub--grid physics, including star formation, radiative cooling and feedback processes. Earlier simulations, which were not succeeding in forming realistic
galaxies due to gas overcooling and weak feedback, routinely obtained overly massive hot gaseous coronae and yielded X--ray luminosities an order of magnitude higher than the contemporary observational constraints \citep{Governato:2004}.

Another unresolved issue in galaxy formation is the so-called \emph{missing baryon problem}. Big Bang Nucleosynthesis and three independent cosmological probes -- cosmic microwave background (CMB, $z\sim10^3$), modeling of the Lyman~$\alpha$ forest ($z\sim3$), hot gas fraction in galaxy clusters ($z\sim 0$) -- yield $\Omega_b \approx 0.04$ \citep[e.g. Planck][]{Collaboration:2013aa}. However, close to our epoch ($z<2$) the number of baryons detected add up to just over half of the number seen at $z>2$ \citep{Nicastro:2005aa, Fukugita:2004aa}, and on the galactic scales it is even less than that, e.g in Milky Way \citep{Dehnen:1998aa,SommerLarsen:2001}, or in M31 \citep{Klypin:2002}. In principle, if a substantial portion of gaseous halo's baryons was 'invisible' in infrared, optical or UV light, but shined and absorbed far--UV and X--ray photons, it would pose a major observational challenge due to the faint and extended nature of coronae around spirals. Successful characterization of such medium would require large--field and highly sensitive instruments, which are not yet available. 

Nevertheless, the existence of such corona has been confirmed to date in various ways. For instance, it has been detected in ultraviolet and X--ray absorption lines of OVI and OVII against 
bright background sources \citep{Sembach:aa, Wang:2005, Peeples:2014, Fang:2015aa}, in OVIII emission \citep{Gupta:2009, Henley:2013aa}, and as excess X--ray emission relative to 
the background \citep{Rasmussen:2009aa,  Anderson:2011aa, Bogdan:2013b, Tumlinson:2011}. Predicted warm--hot ($10^{5-6}$~K) and hot (over $10^6$~K) gas in galactic halos seems 
to be a generic feature, as it is found not only in the Milky Way \citep{Rasmussen:aa, Bregman:2007aa, Miller:2014aa}, but also in extragalactic sources \citep{Anderson:2011aa,Bogdan:2012aa}. 
Thus, there is a clear consensus on the sizable contribution of such gas of halo origin to the baryonic budget \citep{Nicastro:2005aa, Gupta:2012aa}. Nevertheless, mock observations of recent realistic simulations of Milky Way--like galaxies and halos reveal that the detected gas is only a part of the full picture, as some of baryons are missing due to the obscuration by the dense foreground disk. As a consequence, observations of the MW's CGM using various metal lines would generally miss 50\% of CGM gas \citep{Zheng:2015}. However, even with this degree of boosting, the MW's CGM mass would still be low and therefore, it is unclear whether the coronal gas entirely solves the missing baryon problem \citep[e.g.][]{Danforth:2008aa, Miller:2013aa, Werk:2014}.
Additionally, as discussed in \cite{Wu:2001}, if halos contained excessive concentration of hot X-ray emitting gas, this may lead to the over-prediction of the soft X--ray background.


In case hot and warm--hot galactic coronae cannot solve the missing baryon problem while simultaneously satisfying key observational constraints, other factors must come into play. Growing body of observational work shows that galactic outflows are 
ubiquitous \citep[e.g.][]{Martin:2005, Bradshaw:2013, Turner:2014}, therefore baryons could be expelled from the potential wells of galaxies by means of jets, winds and photoionization input from AGN, SNe and OB stars. The resulting distribution of gas would be smoother and more diffuse, reducing the likelihood of detecting this large reservoir of baryons in the Universe.
\cite{Silk:2003} developed an analytic model for such galactic outflows, 
suggesting that the supernova--driven ejection of gas could act before a galaxy assembles \citep[see also][]{Benson:2003aa, Murray:2005aa}. Hence, a galaxy could already be born baryon deficient. 

Early hydrodynamical  simulations of large cosmological volumes of \cite{Dave:2001aa} predicted that the missing baryons could be contained in intergalactic warm--hot and 
hot medium at temperatures $10^5~K<T<10^7~K$. Numerical simulations with effective models of  SNe feedback such as momentum  {and energy}--driven winds have been extensively 
studied by Dav{\'e} and collaborators \citep[e.g.][]{Finlator:2008, Oppenheimer:2008a, Ford:2013}. \cite{Dave:2009} found that such winds eject a large fraction of the baryons 
beyond the potential wells of the galaxies and that this effect is stronger for low--mass halos. According to their findings, the Milky Way--sized halo at $z=0$ is expected to retain 
60$\%$ of baryons within the potential well, and the effect of outflows may extend to as far out as 3 virial radii of the galactic halos. Slightly weaker but still significant baryonic winds are found at $ z > 2.5-5$ in recent hydrodynamical
zoom--in simulations of individual galaxies, in which outflows are generated from the thermal coupling of the SN energy to the surrounding gas \citep{Shen:2012, Shen:2013}.

The goal of this paper is to use three cosmological zoom--ins of Milky Way--sized galaxies from the Eris suite \citep{Guedes:2011aa, Mayer:2012, Shen:2013} to address the abundance of the warm--hot and hot gas in the galactic coronae, along with the larger scale WHIM. In addition to the original Eris, we use runs with the same initial conditions but a much richer inventory of physical processes.  We focus on the properties of the present--day diffuse coronae, as well as on the impact of varying sub--grid physics on the results. Finally, we compare them to the available observational constraints on the density of the Milky Way gas \citep{Gatto:2013aa, Miller:2014aa, Werk:2014} in order to determine whether coronae arising in the simulations may be regarded as realistic. We note that successful recovery of the basic properties of the Milky Way corona is a cornerstone of the forthcoming in--depth study of the coronal assembly (Sokolowska et al., in preparation, hereafter Paper II). 

This paper is outlined as follows. In section 2 we motivate why we investigate this particular set of runs and describe the physics included in the simulations. Section 3 contains our results in the context of Milky Way observables and is followed by the summary in section 4.


\begin{deluxetable}{lccccccccc}
\tablecaption{Parameters of the runs}
\tablenum{1}

\tablehead{\colhead{Run} & \colhead{$z_l$} & \colhead{ M$_{vir}(z_l)$} & \colhead{R$_{vir}(z_l)$} & \colhead{IMF} & \colhead{MC} & \colhead{UVB} & \colhead{$c^*$} & \colhead{$n_{SF}$} & \colhead{ $\epsilon_{SN}$} \\ 
\colhead{} & \colhead{} & \colhead{(10$^{11}$ M$_{\odot}$)} & \colhead{(kpc)} & \colhead{} & \colhead{} & \colhead{} & \colhead{} & \colhead{(cm$^{-3}$)} & \colhead{($10^{51}$erg)} }
\startdata
Eris &  0 &  7.6 &  233 &  K1993 &  low--T &  HM1993 &  0.1 &  5 &  0.8 \\
ELE &  0 &  7.8 &  235 &  K1993 &  low--T & HM1993 &  0.05 &  5 &  0.8 \\
E2k &  0.5 &  6.5 &  170 &  K2001 & all--T & HM2012 &  0.1 &  100 &  1.0 \\
\enddata


\tablecomments{Virial masses and radii of the Milky Way--sized runs were calculated at the {  final redshift of each run} $z_l$. Notation: MC -- metal cooling, $c^*$ -- star formation efficiency, $n_{SF}$-- star formation density threshold, $\epsilon_{SN}$: supernovae efficiency parameter, HM1993: \cite{Haardt:1993aa}, HM 2012: \cite{Haardt:2012aa}, K1993: \cite{Kroupa:1993aa}, K2001: \cite{Kroupa:2001aa}.}
\label{tab:tab1}
\end{deluxetable}


\section{Methods}
\label{sec:methods}
\begin{deluxetable}{lccc}
\tablecaption{Mass of baryons at $z=0.5$}
\tablenum{2}
\tablewidth{320pt}
\tablehead{\colhead{Component} & \colhead{M$_{Eris}$} & \colhead{M$_{ELE}$} & \colhead{M$_{E2k}$} \\ 
\colhead{} & \colhead{(10$^{10}$M$_{\odot}$)} & \colhead{(10$^{10}$M$_{\odot}$)} & \colhead{(10$^{10}$M$_{\odot}$)} } 
\startdata
warm and cold gas  \tiny{($<10^{5}K$)} & 1.20 &  1.27 &  1.22 \\
warm--hot gas \tiny{($10^{5-6}K$)} & 3.02 & 3.98 & 2.43 \\
hot gas \tiny{$(>10^{6}K)$} & 0.63 & 0.72 & 0.69 \\
total gas & 4.85 & 5.97 & 4.34 \\
total stellar &  3.41 & 2.75 & 3.38 \\
\enddata
\tablecomments{Total masses of various components were calculated within virial radii at the corresponding redshift.}
\label{table:masses}
\end{deluxetable}

Here we use three unique high--resolution simulations of spiral galaxies performed with the treeSPH code {\sc Gasoline} \citep{Wadsley:aa} that comprise more than 18~million particles spread over a 90~Mpc box. These runs are all zoom--in simulations of Milky Way--sized galaxies (Table~\ref{tab:tab1}) that are evolved in the full cosmological context in a Wilkinson Microwave Anisotropy Probe 3-year cosmology, $\Omega_M = 0.24$, $\Omega_{\Lambda} = 1-\Omega_M$, $\Omega_b = 0.042$, $H_0 = 73$~km s$^{-1}$Mpc$^{-1}$, $n = 0.96$, $\sigma_8 = 0.76$. The first of the runs, Eris, has been shown to be extremely successful, owing to the recovery of various Milky Way properties \citep{Guedes:2011aa}. Namely, 1)~its rotation curve is in good agreement with observations of blue HB stars in the Milky Way, and it reproduces the kinematic properties of SDSS halo stars \citep{Rashkov:2013}, 2)~stellar mass and the disk scale length are both comparable to the values adopted for our Galaxy, 3)~bulge-to-disk ratio determined by a two-component fit to the i-band surface brightness profile is typical of Sb spirals and Sbc galaxies, 4)~the pulsar dispersion measure gave plausible constraints to the hot halo of the Milky Way \citep[see more in][]{Guedes:2011aa}. 

The two other simulations known as ErisLE \citep[here ELE,][]{Bird:2013aa} and Eris2k (denoted as E2k, Shen et al. in preparation) are the follow--ups of the original Eris. We picked these follow--ups out of a larger set of variants because they made major improvements to the original run and at the same time were able to produce realistic galaxies. We discuss some of these improvements below.

The recipes for star formation and feedback in all runs are the same. Gas particles must be dense (denser than $n_{SF}$ specified in Table~\ref{tab:tab1}) and cool (cooler than $T_{max} = 30 000 K$) to form stars.  Particles which fulfill these requirements are stochastically selected to form stars, based on the commonly used star formation equation $\frac{dM_*}{dt} = c^* \frac{M_{gas}}{t_{dyn}}$, where $M_*$ is mass of stars created, $c^*$ is a constant star formation efficiency factor, $M_{gas}$ is the mass of gas creating the star, and $t_{dyn}$ is the gas dynamical time. Each star particle then represents a population of stars, covering the entire initial mass function. Stars larger than $8~M_{\odot}$ explode as SNII. According to the model of \cite{Stinson:2006aa}, the feedback is purely thermal, as the blastwave shocks convert the kinetic energy of ejecta into thermal energy on scales smaller than simulations can resolve.  Once energy is ejected, particles receiving the energy are prevented from cooling. The reason for this is to mimic two effects: 1)~turbulence in molecular clouds, which inhibits star formation by means of heating up particles, and 2)~a high pressure of the blastwave, as high--temperature gas naturally flows outwards.

The star formation efficiency parameter was lowered in ELE with respect to Eris in order to further improve the observed normalization of the star formation density in local galaxies \citep{Governato:2010}. All other parameters remained the same, as indicated in Table~\ref{tab:tab1} (for a more detailed discussion, see \cite{Bird:2013aa}). In contrast, E2k retains the star formation efficiency of Eris but represents a new generation of runs with a richer inventory of physical processes. These processes include new sub--grid turbulent diffusion prescription for both metals and thermal energy, as well as cooling via metal lines \citep[also see the next paragraph]{Shen:2009aa}. In essence, metals significantly enhance the cooling of the WHIM, while metal diffusion contaminates large amounts of otherwise pristine gas. \cite{Shen:2009aa} have shown that metals mix between winds and surrounding gas before they leave the galaxies, decreasing the metal content in the WHIM and diffuse IGM but increasing it in the galactic halo.

The radiative cooling rates in Eris and ELE (hereafter referred to as the first--generation runs) are obtained by solving non--equilibrium differential equations of primordial gas in the presence of cosmic ionizing background \citep{Wadsley:aa}. Additionally, gas of $T<10^4$~K cools through fine structure and metastable lines of C, N, O, Fe, S and Si \citep{Bromm:2001aa, Mashchenko:2007aa}. E2k run, instead, employs tabulated cooling rates for metal lines at all temperatures. These rates are  calculated with the photoionization code {\sc Cloudy} \citep{cloudy}, which assumes that metals are in ionization equilibrium. This is a good approximation when extragalactic UV radiation is present. 

The strength of feedback depends on the number of produced supernovae, which is in turn governed by the initial mass function. The IMF in Eris and ELE  was based on \cite{Kroupa:1993aa}, while in E2k an updated IMF was used \citep{Kroupa:2001aa}. This difference translates into nearly 3~times more stars in the mass range of $8-40~M_{\odot}$ in the run with the new IMF.  Other significant modifications in the parameters of E2k that might boost feedback are: doubled cooling shut--off time, higher supernovae efficiency parameter $\epsilon_{SN}$ and the higher star formation density threshold $n_{SF}$  (see Table~\ref{tab:tab1}). The latter formally increases the star formation rate by a factor of 4.5, since the star formation rate scales as $\sqrt{n_{SF}}$ \citep[see e.g.][]{Guedes:2011aa, Mayer:2012}. Locally the supernovae rate will therefore increase by the same amount for a given IMF.  This has been shown to improve fidelity of galaxies' properties by means of lowering baryonic densities and stellar masses \citep{Mayer:2012}.

E2k is a run that follows an extensive study of sub--grid parameters. The final parameters were chosen in order to closely match the stellar mass--halo mass relation predicted by abundance matching as a function of redshift
\citep[e.g.][]{Behroozi:2014}. The E2k setup, including the detailed discussion of the stellar component and the structure of the disk, will be presented in the forthcoming paper (Shen et al. in preparation).

In this work we focus on the diffuse ionized gas component exterior to the galactic disk, i.e. the component
traced by gas particles of temperatures \emph{over} $3\cdot10^4$~K. Our classification of the gas phases 
encompassed by such component is inspired by conventions introduced by observers \citep{Putman:2012aa}. 
Specifically, we term \emph{warm gas} the phase comprising particles at $3\cdot10^{4}$~K$<T<10^5$~K, \emph{warm--hot gas} that with temperature in the range $10^5$~K$<T <10^6$~K, and  \emph{hot gas} the phase at $T > 10^6$~K.


\section{Results}
\label{sec:results}

Given that the E2k has not yet been run to $z=0$,  we compare all three runs at the last common 
redshift, i.e. at $z=0.5$. There is no major merger after $z \sim 2$ in any of the runs and no major satellite
is accreted below redshift $z \sim 0.5$ \citep{Pillepich:2015}. As a consequence, the thermodynamical state of the gas at $z=0.5$ should not differ appreciably from that at $z=0$, and we have verified this in {  the }case of Eris and ELE.

\begin{figure}[h!]
\centering
\figurenum{1}
\plotone{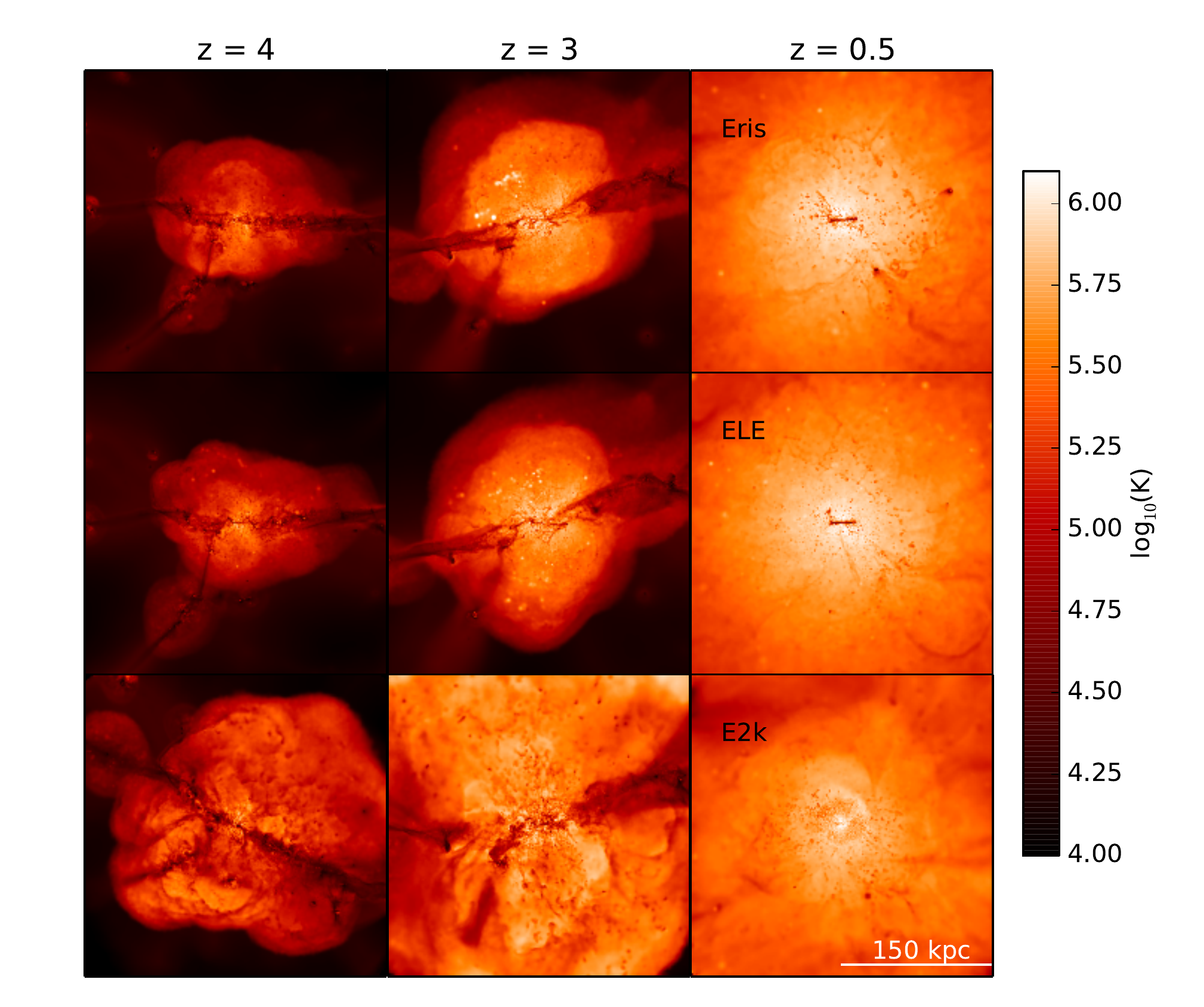}
\caption{Temperature maps of the three representative runs: Eris (top row), ELE(middle) and E2k (bottom row), scale is in the physical units (kpc). Snapshots correspond to three example evolutionary time steps.  Virial radii in all three runs are approximately $r_{vir}=(30,50,$170$)$~kpc for the corresponding redshifts $z=(4,3,$0.5$)$. Note the white region at $z=~0.5$ -- it is a signature of the spherically--shaped hot coronae.}
\label{fig:fig1}
\end{figure}

The most significant similarity of all runs lies in the amount of gas residing in the warm--hot phase at $z=0.5$, 
together with the hot gas mass (see Table~\ref{table:masses}).  Warm--hot and hot gas together are of the 
order of stellar mass, and constitute as much as 72--80\% of the total gas content, 
substantially exceeding the contribution of warm and cold (i.e. $T<3\cdot 10^4$~K) gas. 
This has important implications for the missing baryon problem (see Section \ref{sec:mbp}). 

Fig.~\ref{fig:fig1} illustrates the evolution of the gaseous halo across three representative times 
for the first--generation runs, and for E2k. Filamentary feeding occurs very early on (at $z=4$) as IGM loads mass into the neighborhood of a protogalaxy.   The effect of feedback associated with the majestic expanding bubble seen between redshifts $z=4$ and $z=3$ facilitates redistribution of energy and mass far beyond the virial radius. With the passage of time, the halo starts resembling an onion--like structure -- a corona of hot gas is embedded in a warm--hot bubble. As argued above, significantly more supernovae explode in E2k relative to the first--generation runs, which explains the differences in temperature between the top and bottom panels of Fig.~\ref{fig:fig1}. Also, in E2k the stronger feedback can push baryonic matter much farther out from the potential well of the galactic halo with respect to the first--generation runs.

All runs feature generic accumulation of warm--hot and hot gas component through the combination of feedback and shock--heating after the last major merger ($z<3$, see Figure~\ref{fig:fig1}). The specific role of these distinct processes in 
building the galactic halo will be 
addressed in Paper II.
In the remainder of this section we will present the late--time properties of the corona and focus on how they compare with the latest observational constraints on the diffuse gaseous halo of the Milky Way. 

\subsection{Mass budget of the corona}
\label{sec:mbp}

\begin{figure}[h!]
\centering
\figurenum{2}
\label{fig:fig2}
\plotone{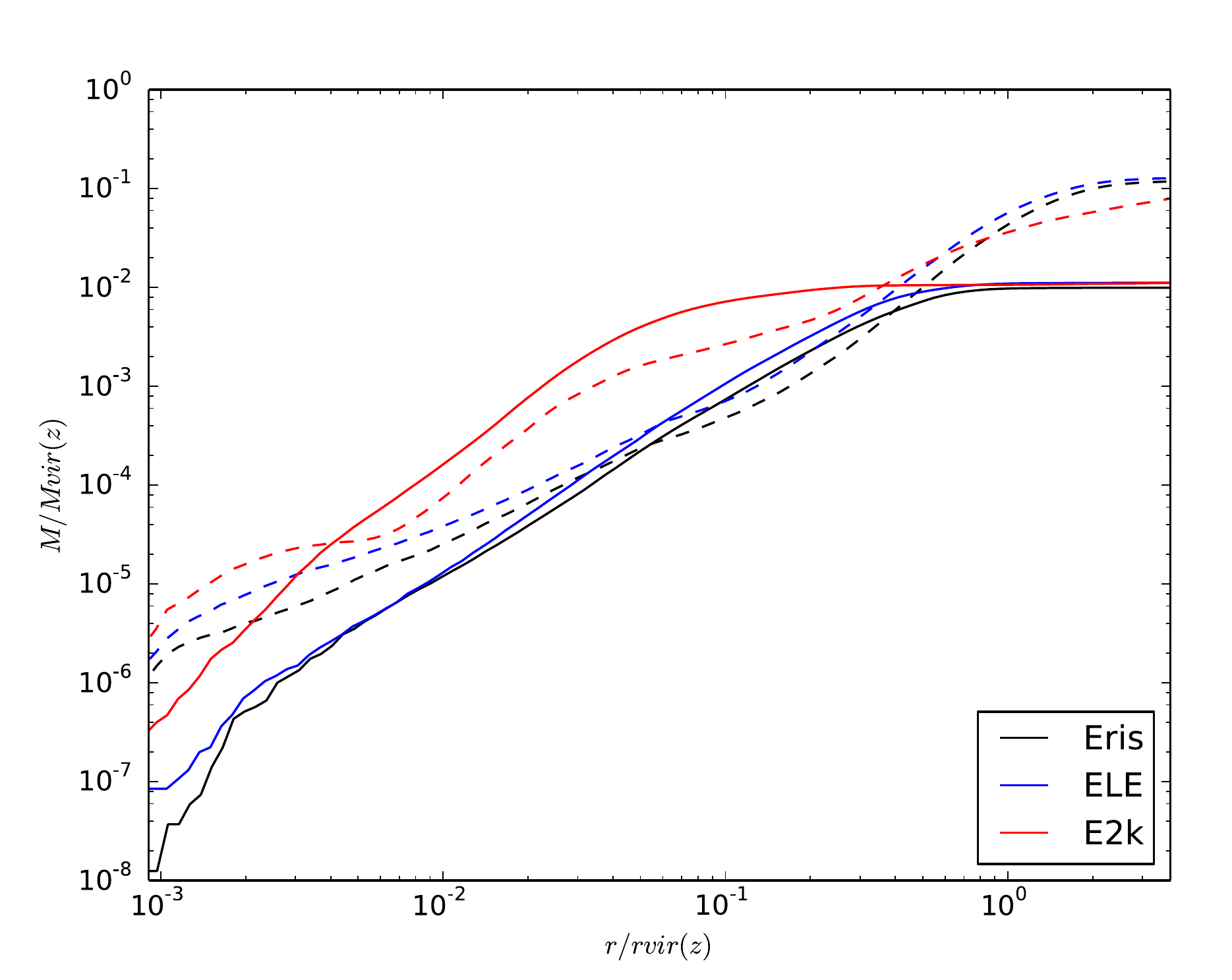}
\caption{Radial profile of the cumulative mass of warm--hot gas (dashed) and hot gas (solid) at $z=0.5$ for three independent runs. Both mass and radial bins are normalized to the exact virial mass  $M_{vir}=~(6.5, 6.7, 6.5)~10^{11}~M_{\odot}$ and virial radius $r_{vir}=~(168.5, 170.4, 169.7)$~kpc of Eris, ELE, E2k respectively. }
\end{figure}

In Fig. \ref{fig:fig2}, we present the cumulative radial distribution of warm--hot and hot gas mass normalized to the virial mass at $z=~0.5$.  Although E2k clearly stands out (red lines), Eris and ELE exhibit similar distribution of both components, with a slight overabundance of gas in ELE with respect to Eris, likely due to the lower efficiency in turning gas into stars. 

All three runs share a common attribute -- hot gas attains a fraction of the virial mass of order 1\%, and the location of this maximum indicates that hot coronae are enclosed within $\sim$100~kpc at $z=0.5$ (see also Fig.~\ref{fig:fig1}). At $z=0$ this translates into hot phase gas enclosed within $\sim$140~kpc, hence much below the virial radius of the galaxy. It is the warm--hot medium that extends to the virial radius ($r_{vir}\sim$ 240~kpc at $z=0$) and beyond.  We note that the size of the Milky Way gaseous halo is still a huge uncertainty -- according to studies of Local Group by \cite{Nicastro:aa} and \cite{Rasmussen:aa}, the corona of hot gas may extend to 1~Mpc or over 140~kpc respectively; \cite{Bregman:2007aa} favor the range of 15--100~kpc, while \cite{Gupta:2012aa} mention over 100~kpc.

Despite these similarities, we find considerably higher proportion of hot and warm--hot gas in the inner part of 
the E2k halo relative to the first generation runs, which signals higher efficiency of feedback. We would then 
expect this to be true even at larger radii, while instead close to the virial radius the
relative proportions of hot gas are the same as in the other runs, and of warm--hot gas the lowest of all, as shown in Figure~\ref{fig:fig2}.

\begin{deluxetable}{lccc}
\tablecaption{Baryonic fraction $f_b$ at $z=0.5$}
\tablenum{3}
\tablewidth{250pt}
\tablehead{\colhead{Run} & \colhead{$f_b$} & \colhead{$f_b$} & \colhead{$f_b$} \\ 
\colhead{} & \colhead{($<r_{vir}$)} & \colhead{($<2~r_{vir}$)} & \colhead{($<3~r_{vir}$)} } 

\startdata
Eris & 0.124 (71\%) &  0.145 (83\%) &  0.160 (91\%) \\
ELE & 0.130 (74\%) &  0.149 (85\%) &  0.162 (92\%) \\
E2k & 0.117 (67\%) & 0.115 (66\%) & 0.129 (74\%) \\
\enddata
\tablecomments{Baryonic fraction $f_b$ at $z=0.5$ was calculated for three spherical regions characterized by a multiple of a virial radius.  Numbers in brackets correspond to the percentage of the recovered baryons with respect to the predictions of the Wilkinson Microwave Anisotropy Probe 3-year cosmology, i.e. $f_b = 0.175$. The same cosmological parameters were used to generate initial conditions of all 3 runs.}
\label{tab:tab2} 
\end{deluxetable}
\clearpage

Differences in the cooling schemes provide one viable explanation. In this scenario the enhanced metal cooling in E2k transforms the hot phase into the warm--hot phase more efficiently. This would be more evident further away from the center, as the competitive effect of heating by feedback becomes progressively weaker.  Similarly, metal cooling below $10^6 K$ could replenish the warm and cold gas phase more efficiently at the expense of the warm--hot gas.  This effect should be even stronger, given the higher gas densities implied in this case, as well as the fact that the cooling curve peaks near
few times $10^5$~K.

However, the numbers presented in Table~\ref{table:masses} disprove that scenario, as total mass of warm--hot gas of E2k is the lowest of all, and we observe no significant difference in the mass of warm and cold gas relative to the other runs. We thus conclude that some gas in E2k is heated to higher temperatures by stronger feedback and simply leaves the halo. The stronger baryonic outflow is indeed confirmed by the clear difference in the baryon fraction within $R_{vir}$ between E2k and the first generation runs shown in Table~\ref{tab:tab2}.

The extent, as well as the total mass fraction of the warm--hot medium, are of particular importance for the missing baryon problem. In three independent runs the warm--hot medium exceeds the hot phase in mass by a factor of 4--5. Moreover, as indicated in Table~\ref{table:masses}, warm--hot gas itself constitutes the majority of the global gaseous budget of each galaxy. When the warm--hot component is removed from the calculation, the baryonic fraction amounts to 42--47\% of the universal value at $z=0.5$, in fact a fraction that is very close to the one observed at $z<2$ \citep{Nicastro:2005aa}. 

With the warm--hot gas in the picture, the baryonic fraction is much higher than the one quoted above, though lower than the one inferred from the WMAP3 cosmology (see the first row of Table \ref{tab:tab2}). We recover about 67\%--74\% of the cosmic baryon budget within the virial radius. We are roughly consistent with the work of \cite{Dave:2009}, who finds that the baryonic fraction in the present--day Milky Way--sized halos ($\sim 10^{12} M_{\odot}$) drops to about 60\% of the cosmic value. This agreement is striking, given the very different sub--grid recipes for feedback, as well as more than an order--of--magnitude difference in mass and spatial resolution of the simulations. We note, however, that the different feedback recipe that we adopt is likely crucial in obtaining realistic properties at the level of the galactic disk and below \citep[e.g.][]{Bird:2013aa, Guedes:2011aa}.

Next, we investigated the efficiency of feedback--driven galactic winds in removing baryons within the different
runs. We therefore calculated baryonic fractions within 2 and 3 virial radii (Table~\ref{tab:tab2}) and found that they increase substantially beyond the galactic halos, which reveals that galactic halos are surrounded by large reservoirs of baryons. Note that there is a correlation between strength of feedback and the value of the baryonic fraction, that is the stronger the feedback, the lower the mass of baryons encompassed. E2k attains the lowest value of all, while ELE as the run of the lowest stellar mass (hence the weakest feedback) exhibits the highest $f_b$. 

We thus conclude that galactic outflows triggered by supernova feedback are responsible for the \emph{missing} of a significant fraction of the baryons from the galactic halos, and these baryons can be found mostly in warm--hot intergalactic medium that extends even further than 3~virial radii (Fig.~\ref{fig:fig2}). The fact that the first--generation runs assembled up to 92\% of cosmic 
mean in baryons within 3 virial radii suggests that the missing of the baryons is likely a scale--dependent effect. Once the baryonic census is undertaken by designing observations of warm--hot and
hot gas extending to sufficiently large radii around virialized structures, the missing baryon problem might be ultimately explained.

\subsection{Density and entropy of the galactic corona}
\label{sec:sec1}
\begin{figure}[h!]
\centering
\figurenum{3}
\label{fig:fig3}
\plotone{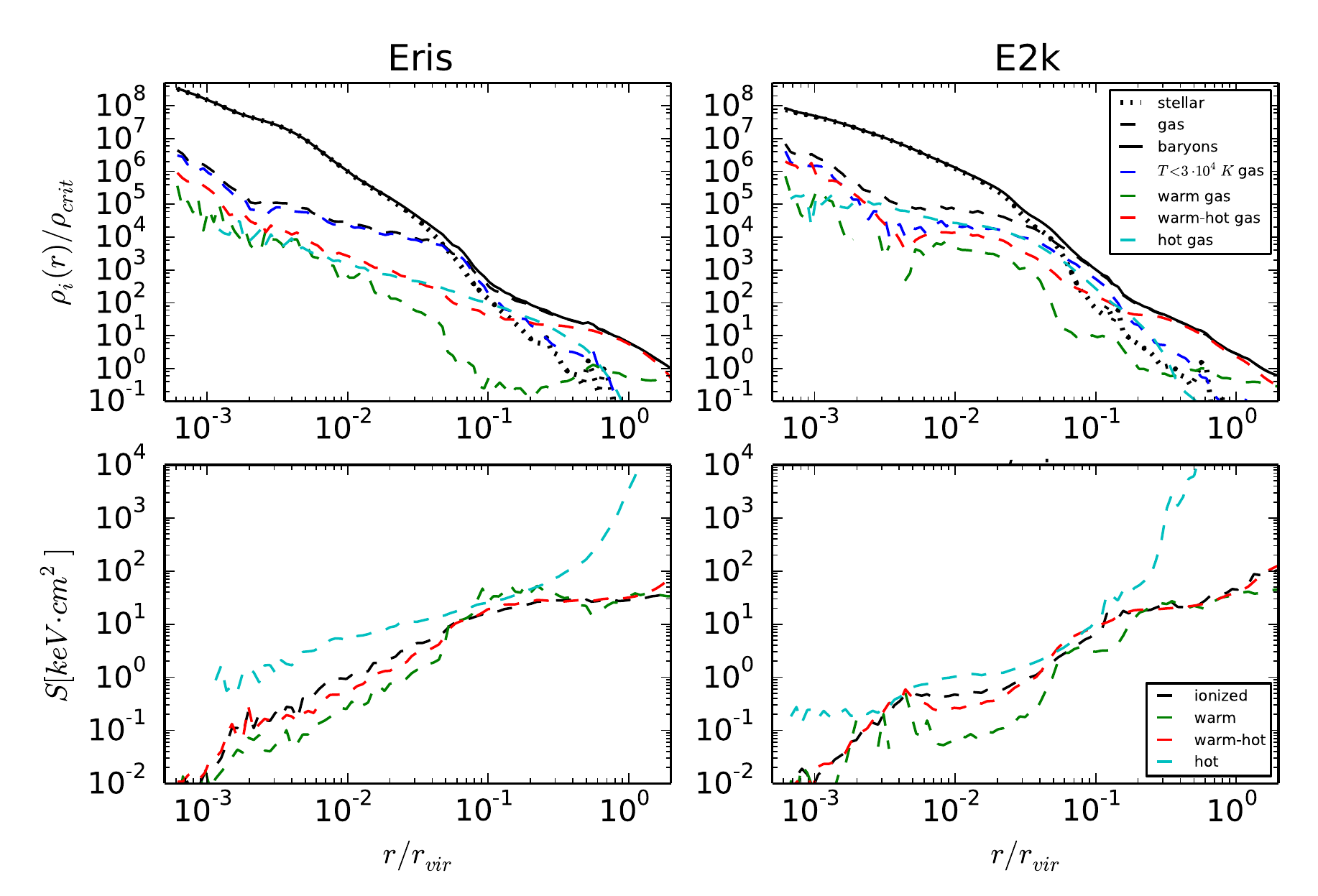}
\caption{Top row: mass density distributions of stars (dotted black), gas (dashed black) and baryons (solid black) at $z=0.5$. Particular constituents of gas are presented with colored dashed lines (see legend).  Although we concentrate on the ionized gas in this work, we included mass density of cold gas of $T<3\cdot 10^4$~K (blue dashed line) as well. Bottom row: entropy of ionized gas phases that incorporate an additional \emph{ionized} term to denote all ionized particles.}
\end{figure}

In this section we scrutinize various components: baryons, stars and gas to pinpoint the differences in thermodynamics of the runs that nevertheless lead to similarities content--wise. To display the collective effect of thermodynamical processes, we use mass density distributions of gas phases mentioned in section~\ref{sec:methods} normalized to the critical density of the Universe at $z=0.5$, as well as radial entropy profiles. The proxy for entropy that we assumed here is commonly used in studies of clusters \citep[e.g.][]{Babul:2002aa, Voit:2005aa, McCarthy:2007aa}, and is defined as $S = k_B T/n_e^{2/3}$. We note that the distributions of the first--generation runs are seemingly alike, therefore we chose Eris to be  representative of these in comparison with E2k (see Fig.\ref{fig:fig3}).

Certain similarities are visible at the first glance, such as the dominance of the stellar content in the inner part where the galactic disk resides, along with the convergence of total gaseous and baryonic distribution outside the disk (see Fig. \ref{fig:fig3}, top row). E2k however exhibits higher stellar concentration in the very center, and so higher concentration of baryons.  Nevertheless, a steeper stellar density slope of E2k cancels out this difference, as the total stellar mass for both runs at $z=~0.5$ amounts to $3.4\cdot10^{10} M_{\odot}$. 

Sub--grid physics behind the two generations of runs shapes the intrinsic properties of the galactic disks. As a consequence, it governs mass distributions of gas phases, e.g. leading to the closer confinement of hot gas in E2k. Cold gas with temperatures below $3\cdot 10^4K$ dominates the first--generation runs in the center, whereas E2k exhibits a comparable contribution of cold, hot and warm--hot gas from $\sim 350$~pc outward, probably facilitated by the diffusion of metals and thermal energy. Outside a region inhabited by a galactic disk, all simulations appear to have been mildly affected by the tremendous differences in the disk, such as very high metallicity of the disk gas (Shen et al. in prep.). Beyond 0.2~r$_{vir}$, warm--hot medium forms the majority of baryons, while cold and warm components are allocated comparably. 

The warm medium exhibits a characteristic drop down in its mass density distribution (green dashed line in Fig.~\ref{fig:fig3}) localized at roughly $0.4~r_{vir}$ and continues up to a point where hot and warm--hot gas eventually begin to saturate the cumulative distribution of their mass. A resulting feature, the \emph{density dip}, underlines an 
on--going thermodynamical process that triggers a local decrement of warm gas. When combined with the temperature maps, which show lack of warm gas in the vicinity of the dip starting at $z=1$ (visible later as the hot region in Fig.~\ref{fig:fig1}), a 
causal connection with some event implied in the assembly of the corona is strongly suggested. Further time--evolution analysis is necessary to address the origin of such feature and will be undertaken in Paper II.

Cooling and heating mechanisms leave their imprint in the entropy distribution, therefore entropy itself carries information about the thermodynamical history of the gaseous halo. In the bottom row of Fig. \ref{fig:fig3} we present a preliminary result of the analysis of the entropy of gaseous halo constituents.  The entropy profile of warm gas reflects specific features of the density distribution, due to nearly constant temperature profile of that phase. Hence, the density dip manifests itself as an \emph{entropy bump} in Fig. \ref{fig:fig3} (bottom row), indicating the possible deposition of heat in that region. 

We note that entropy profiles presented here were shown for the phases as we defined them since the beginning of the paper. In principle,
useful physical insight could be gained by re--defining phases based on eventual regularity in their entropy profiles, for example
highlighting possible near--equilibrium configurations. Such an approach would be physics--driven rather than observationally--driven,
and will be adopted in Paper II.
Nevertheless, the entropy of the diffuse component is free of potential bias, and appears to be consistent with 
predictions for the hot X--ray emitting gas in clusters, as discussed in the next paragraphs.


Unlike the case of the warm phase, the temperature profiles of warm--hot, hot and all ionized gas are far more complex than simple power laws.
Reflecting the different strength of radiative cooling and heating by feedback, the shape of entropy profiles of such phases
differs substantially between the runs. For example, the entropy distribution of hot gas in Eris has a shallower slope than any other component up 
to $0.1~r_{vir}$(cyan line), whereas in E2k flattens towards the center of the halo in a sort of "entropy core", resembling that
found in some galaxy clusters. In both cases entropy rises dramatically beyond that radius, though at a different rate.
Therefore entropy, as widely shown in the literature on galaxy cluster simulations, is a sensitive diagnostic 
of sub--grid physics.

Despite differences, we identify a generic feature -- the entropy distributions of the warm--hot component (red line) and the entire ionized gas (black line) are nearly identical 
for a given run. This follows from the fact that warm--hot medium constitutes the majority of the overall gas mass. Moreover, the specific entropy of the ionized gas is leveled in both cases between 0.1-- 1~r$_{vir}$ and reaches $40-50$~keV~cm$^2$ at the virial radius. We compared with \cite{McCarthy:2007aa} that this is consistent with the value of a characteristic entropy at large radii expected for a halo of mass of order $10^{12}M_{\odot}$. In case of E2k, the entropy rises further beyond the virial radius, which is due to the stronger effect of feedback, as discussed in section~\ref{sec:mbp}. 

\subsection{X--ray luminosity}
One of the hot halo thermal properties that can be readily compared with the observations is the X--ray luminosity. We compute the $0.5-2.0$~keV band X--ray luminosity of the coronal gas using the radiative rates of Astrophysical Plasma Emission Code3 (APEC) from \cite{APEC}.  APEC assumes the optically thin gas in collisional ionization equilibrium. The final value is a sum over the X--ray luminosities of individual gas particles within a distance $r_{vir}$ of the halo center, taking into account contributions to line and continuum emission associated with each of the individually tracked elements (iron, oxygen, hydrogen and helium). Total X--ray luminosity is defined as
\begin{equation}
L_X = \sum_k n_{e,k} n_{i,k} \Lambda_k V_k ,
\label{eq:Lx}
\end{equation}
with $n_{e,k}$ the number density of electrons, $n_{i,k}$ the number density of ions,  $V_k = m_{k}/\rho_{k}$ the volume  and $\Lambda_k$ the cooling rate of an individual gas particle $k$. The resultant X--ray luminosities for Eris, ELE and E2k are: $4\times 10^{39}$~erg~s$^{-1}$, $10^{40}$~erg~s$^{-1}$ and $3\times 10^{42}$~erg~s$^{-1}$, respectively.

\cite{Snowden:1997} modeled a bulge of hot gas surrounding the Galactic center, based on the maps of the soft X-ray background from the ROSAT all-sky survey. They assumed a cylindrical geometry with an exponential fall--off of density. At the radial extent of $\sim 5.6$~kpc and a scale height $\sim 1.9$~kpc, the total luminosity in the 0.5--2~keV band that they find is ~$\sim 2\cdot 10^{39} \text{erg s}^{-1}$. Other constraints follow from \cite{Wang:1998}, who characterized the soft X--ray background of the hot polytropic corona of the Galaxy, using all--sky ROSAT emission. In their model, Galactic corona is assumed to be axisymmetric and quasi--hydrostatic. The integrated luminosity in the 0.5--2~keV range that they find is $\sim 3\cdot 10^{39} \text{erg s}^{-1}$. More recently, \cite{Miller:2014aa} reported the X--ray luminosity of the Milky Way corona in that band in the metallicity--dependent range of $0.8-5.0 \cdot 10^{39} \text{erg s}^{-1}$. They confirm the luminosity found by \cite{Wang:1998} for the $Z=0.3~Z_{\odot}$ corona, which lies in--between the mean metallicities of the gaseous galactic halos of Eris ($Z \sim 0.1~Z_{\odot}$) and Eris2k ($Z \sim 0.6~Z_{\odot}$), and is comparable with the mean metallicities of hot gas beyond 20~kpc (in our simulations, $Z \in (0.22, 0.26)Z_{\odot}$). We conclude that {  the X--ray luminosities of the coronae of Eris and ELE are consistent with the constraints inferred for the Milky Way corona, whereas E2k attains $L_X$ that is 100 times higher than these constraints.}We discuss this difference in section~\ref{sec:discussion}.

Numerous works reported measurements of X--ray luminosities of the coronae of other spiral galaxies, which cover the broad range of luminosities found in our runs. For example, \cite{Bogdan:2012aa} find luminosities of order $\sim 2 \cdot 10^{40} \text{erg s}^{-1}$ in the radial bin (0.05--0.15)~r$_{vir}$ of massive halos ($\sim 10^{13}M_{\odot}$) of highly star--forming ($15~M_{\odot}yr^{-1}$) spirals.  \cite{Bogdan:2015aa} attempted to detect X--ray coronae around lower--mass spirals and derived the 3$\sigma$ limits on their characteristics. The smallest objects in their sample -- NGC~1097, NGC~5170 -- have estimated SFRs of $(5.8, 0.4)~M_{\odot}\text{yr}^{-1}$ and radii enclosing 200 times the Universe's critical density of $r_{200}=(238,269)$~kpc, which roughly corresponds to the enclosed halo masses of $M_{200}\simeq(1.6, 3.7)\cdot 10^{12}M_{\odot}$. The measurements yield the upper limits on their X--ray luminosities in the 0.5--2~keV band of order $(1.5,3.0)\cdot 10^{39} \text{erg s}^{-1}$, respectively. Among other works, \cite{Strickland:2004a} measured 0.3--2~keV band luminosity of a Milky Way--like, $L^*$ galaxy NGC~891 and obtained $L_X=1.3\cdot10^{39}\text{erg s}^{-1}$, while \cite{Tullmann:2006} reported gas luminosities in the same band of $9\cdot10^{38}\text{erg s}^{-1}$ and $3\cdot10^{38}\text{erg s}^{-1}$ for two non--starburst, "normal" SBc, Sc galaxies NGC~3044 and NGC~4634.

\subsection{Comparison with the Milky Way corona}
\label{sec:sec3}
Understanding the true nature of galactic coronae is a crucial part of the galaxy formation theory. In order to assess our models, we need to determine whether the structure of the simulated halos agrees with the available observational limits. In what follows, we describe three most recent works focused on constraining the number density of the Milky Way halo gas, as well as discuss how our three runs compare with observed Milky Way--like galaxies.

One of the most recent successful detections of hot gas in the Milky Way is based on the measurements of column densities of electrons derived from OVII absorption lines \citep[see][]{Sembach:aa, Bregman:2007aa, Miller:2013aa}. \cite{Miller:2013aa} find their absorption measurements to be well described by a spherical $\beta$ model
\begin{equation} n_e(r) = n_o[1+(r/r_c)^2]^{-3\beta/2},
\end{equation}
hereafter denoted as M$\&$B model. Due to low number of targets passing near the Galactic center, the degeneracy between central density $n_o$ and the core radius 
$r_c$ occurs. Therefore in the regime $r>>r_c$ (typically $r_c\lesssim 1$~kpc) the function may be approximated by $n_e(r) \approx n_o (r_c/r)^{3\beta}$ with $n_o r_c^{3\beta}= 0.013^{+0.016}_{-0.010}$~cm$^{-3}$~kpc$^{3\beta}$ being the core density, and $\beta = 0.56^{+0.10}_{-0.12}$ the slope at large radii, parameters found for the optically thin medium.  This asymptotic profile is what we use in this work as the reference data. 

\cite{Miller:2013aa} derive the electron number density profile, $n_e(r)$, from the electron column density $N_e$. As shown in their work, the latter is related to the metallicity $Z$ and OVII ionization fraction $f$ of the gas as $ N_e \propto f^{-1} Z^{-1}$. Metallicity and ionization fractions are assumed to be constant in the M$\&$B model, i.e. $Z=Z_{\odot}$ and $f = 0.5$ throughout the entire gaseous halo. This is an oversimplification, as in reality both of these quantities have radial distributions. We therefore modify the above M\&B model by replacing the constant factors with the functions of the distance $Z=Z(r)$ and $f=f(r)$ that are inferred from the simulations. We explain this procedure in more detail below. 

The distances between an observer at $r=0$ and a target $r_i$ are in fact equivalent to the path lengths $l_i$. Following the M$\&$B model, since the number of targets is limited, we assume both the spherical symmetry and the constant mean number density $\bar{n}_{e,i}$ between these targets \citep[see also][]{Bregman:2007aa}. Then, for example, equation~(3) holds in case of the first target in a sample, while the number density towards the next target can be obtained from the equation~(4).

\begin{eqnarray}
N_{e,1} = \int_0^{l_1} n_e dr = \bar{n}_{e,1}l_1 \\
N_{e,2} = \int_0^{l_{2}} n_e dr = \int_0^{l_1} n_e dr + \int_{l_1}^{l_{2}} n_e dr =\bar{n}_{e,1}l_{1} + \bar{n}_{e,2}(l_2 - l_1)
\end{eqnarray}
In general, with known path lengths $l_i$ and electron column densities $N_{e,i}$,  the mean number density of each spherical shell $\bar{n}_{e,i}$ can be obtained recursively, according to
\begin{equation}
N_{e,i} = \int_0^{l_i} n_e(r) dr = \sum_i \bar{n}_{e,i}(l_{i+1}-l_i).
\end{equation}
Combined information about the number densities between available targets allows for the recovery of the approximate distribution $n_e(r).$

Total electron column density $N_e$ stems from the measured column density of oxygen ions, $N_{OVII}$, hence the relationship $N_e \propto f^{-1} Z^{-1}$ cited above. We thus apply the correction for $f(r)$ and $Z(r)$ in two steps: 1)~we remove the constant $Z$ and $f$, which is a straightforward multiplication of equation~(5) for all $i$, 2)~we correct each slab of volume between the targets individually for its mean metallicity $\bar{Z}_i$ and mean ionization fraction of OVII $\bar{f}_i$ computed in the simulation. In this procedure the re--calibration factor $\bar{f}_i^{-1} \bar{Z}_i^{-1}$ is thus simply a scaling factor applied to the integrals that varies shell--by--shell. We re--calibrate each run separately in this way. 

As our simulations do not trace OVII, we use the photoionization code {\sc Cloudy} to compute, in post--processing, the ionization fractions of oxygen for each SPH particle. The code assumes an optically thin slab of gas at the density, temperature and metallicity of the SPH particle, irradiated under an isotropic cosmological UV background \citep{Haardt:2012aa}. The effect of local radiation is not included in the calculation.

\begin{figure}[h!]
\centering
\figurenum{4}
\label{fig:fig4}
\plotone{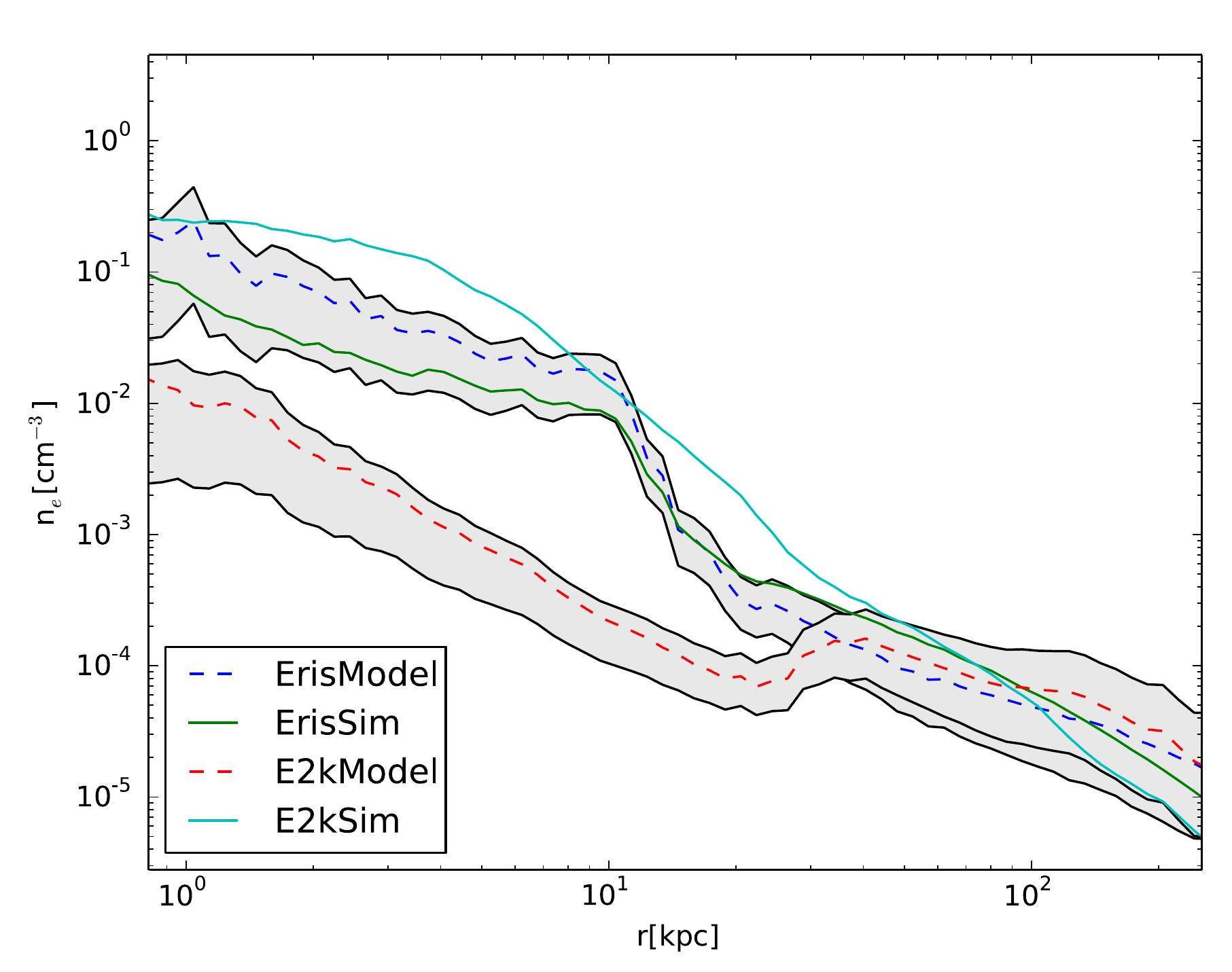}
\caption{Total electron number density of the Milky Way inferred from corrected M$\&$B models -- ErisModel, E2kModel (dashed lines) -- set against our MW-like runs ErisSim, E2kSim (solid lines). Both runs are presented at the latest redshift this time, i.e. Eris is plotted at $z=0$, while E2k at $z=0.5$. The grey shaded regions represent expected error bars on OVII absorption measurements taken from \cite{Miller:2014aa}. }
\end{figure}

In Figure \ref{fig:fig4}, we present a comparison of the total electron number density distributions inferred from our simulations (Eris at $z=0$ and E2k at $z=0.5$) and the two corrected M$\&$B model curves. The sharp drop-off in the electron density profile at about 10~kpc likely signals the extent of the star--forming disk. Clearly, the Eris run agrees with the predicted density distribution for the Milky Way even in the central part, while the 
E2k run fails to recover the inner 30~kpc. 

Different metallicity and ionization fraction profiles of the runs trigger the discrepancy between the two theoretical curves in Fig.~\ref{fig:fig4} in this region (dashed lines).  As seen in Fig.~\ref{fig:fig3}, E2k has a higher density of hot gas in the center than the first generation runs, which goes hand in hand with the higher ionization fraction in the vicinity of the disk (roughly a factor of 6). Additionally, metals are consistently more abundant within the radial 30~kpc -- we refer to this in section~\ref{sec:discussion}. Both, the higher ionization fraction and metallicity are causally linked to the more powerful SN feedback in E2k. 

Despite the differences in the inner part of the corona, this simulation and the derived corrected M\&B model results are in excellent agreement beyond the immediate vicinity of the disk (within the error bars). Additionally, \cite{Anderson:2010} report an average electron number density between the Sun and LMC (50~kpc) of  $5\times 10^{-4} \text{ cm}^{-3}$. In Eris and E2k, the mean electron number densities are $3 \times 10^{-4}$ and $5\times 10^{-4} \text{ cm}^{-3}$, respectively.

\begin{figure}[h!]
\centering
\figurenum{5}
\label{fig:fig5}
\plotone{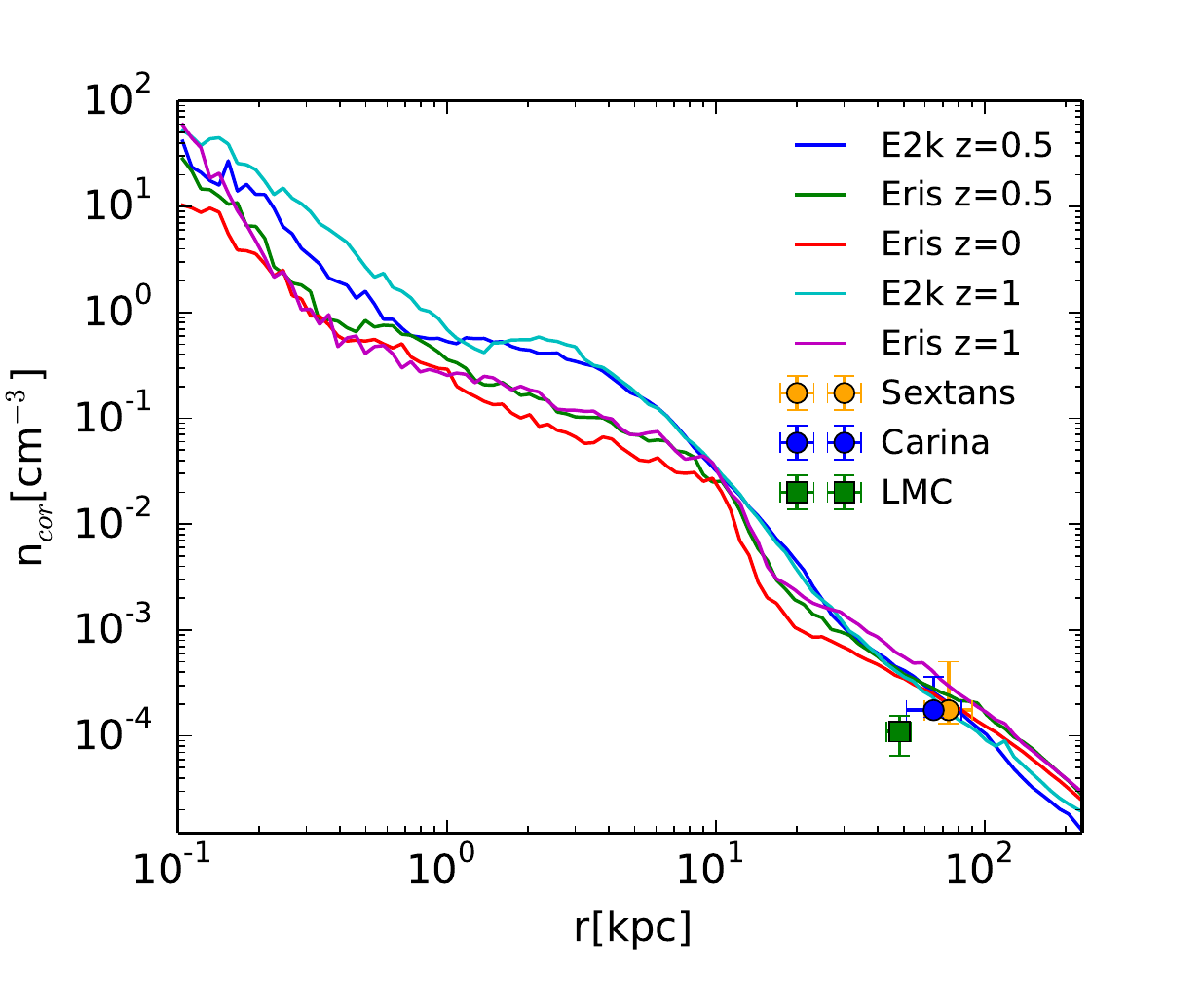}
\caption{Spherically--averaged gas number density profiles in Eris and E2k after $z=1$. The two data points for satellites Sextans and Carina were taken from the Table~6 of \cite{Gatto:2013aa}{ , and the data point for the LMC is from \cite{Salem:2015}}.}
\end{figure}

Another constraint follows from \cite{Gatto:2013aa}, who deduced upper and lower limits on the number density of the Milky Way gaseous halo from the argument of ram--pressure gas stripping of dwarf satellites Sextans and Carina. We use their results as yet another test for the similarity of the gas in simulations and in the Milky Way and we set them in Fig.~\ref{fig:fig5}, where we present coronal number density distributions at different redshifts. $n_{cor}$ incorporates all particles: neutral atoms (in the part where cold gas resides), ions and electrons. The density of gas necessary to efficiently remove gas from the Milky Way dwarfs (shown by the data points with the corresponding error bars) is in excellent agreement with the  
density of gas  in all simulations, as long as satellites plunge in to at least $50-80$ kpc, which is expected based
on the estimates for the orbital parameters of most dSphs. 

Recently, \cite{Salem:2015} modeled the ram pressure stripping of another satellite of Milky Way, LMC, both analytically and numerically. They obtained a lower value than \cite{Gatto:2013aa}, namely $1.1\times 10^{-4}$~cm$^{-3}$ at $\sim 50$~kpc {,} which we also show in Fig.~\ref{fig:fig5}.  {  This density is also lower than the gas density in all three of our simulations, by approximately a factor of 3.   We will discuss the scatter in halo density inferred from the observations as well as the apparent tension between the \cite{Salem:2015} results and our simulations further in section \ref{sec:discussion}. }

\subsection{$L \sim L^*$ galaxies}
In sections~\ref{sec:results} and \ref{sec:mbp} we discussed the relevance of warm--hot and hot phases in accounting for the missing baryons, and in Table~\ref{table:masses} we compared their masses to the other baryonic components.  
Here, we perform another comparison with observations that addresses the weight of the different phases as a fraction of the cosmic baryonic budget. The data is inferred from the recent analysis of a large sample of bright galaxies surveyed by COS \citep[][see also references therein]{Werk:2014}.  COS investigated gaseous halos of 44 $L \sim L^*$ low--redshift galaxies extending out to 160~kpc away 
from the center -- in fact a value that corresponds to the virial radius of our simulated galaxies at $z=0.5$. While the direct focus of the survey was to use
a set of low ionization--state tracers to measure the mass of the gas at $T < 2 \times 10^5$ K, it combined input from other surveys of tracers of hotter
gas to provide a fairly complete census.
\cite{Werk:2014} assumed the cosmic baryon fraction of $0.17~M_{vir}$, therefore, to be consistent, we change their normalization from 0.17 to 0.175 which is a cosmic baryonic fraction in the cosmological framework of the simulations.

We calculated that the {  combined} warm--hot and hot components in the simulations represent 32$\%$, 40$\%$ and 27$\%$ of the baryonic budget in Eris, ELE and E2k respectively, which agrees with the predicted range of 7--45$\%$ in the COS Halos, assuming the cosmic baryon fraction in the WMAP3 cosmology. Given the large span between the lower and the upper limit on the mass of the components, we treat 
this comparison with caution. The tendency of our results to lie at the upper end of the values inferred by \cite{Werk:2014} reflects the fact that their
typical galaxy has a higher fraction of warm and cold gas relative to what we find in our runs. 

Among caveats to take into account in this comparison is that the mean estimated halo mass in the COS sample,  $10^{12.2}M_{\odot}$, is higher than the virial mass of our galaxies. In particular, the fiducial COS halo
would lie above the transition mass scale between cold and hot mode dominance in the accretion flow, which could have an impact on gas thermodynamics. Furthermore, the stellar mass of the galaxies in the COS sample, which enters in the computation of baryonic mass fractions, is not determined observationally but 
simply deduced from the abundance matching. Finally, COS--Halos sample the redshift $z=0.2$, while the simulation outputs are taken at $z=0.5$. Notwithstanding these caveats, this comparison strengthens the relevance of our results, since it gives them statistical significance as opposed to restricting the comparison only to the Milky Way.  { As pointed out by \cite{Zheng:2015}, there are discrepancies between the mass budgets of MW's and COS halos' CGM, opening up a possibility that either the Milky Way is an outlier, or that the observations are still incomplete. Hence, this complementary comparison going beyond the Milky Way data is of a tremendous value for our results.}

\section{Concluding remarks}
\label{sec:discussion}
In this paper we presented three simulations of MW--like galaxies. It is important to emphasize that these simulations are high--resolution zoom--ins carried out in a fully cosmological context. These simulations produce disk--dominated galaxies that are devoid of overcooling problem or an overly massive stellar component, as thoroughly discussed in previous papers on the Eris suite of simulations.  The differences in gas distributions between Eris and ELE were minor, therefore we chose Eris to represent the group regarded here as the \emph{first--generation} runs.  E2k run differs from that group, as it comprises richer inventory of physical processes.  As will be discussed in Shen et al. (in prep.), E2k yields a lower stellar mass at high redshift relative to
first generation runs, showing a better agreement with abundance matching. At the same time, it hosts a more turbulent, thicker disk compared to that of Eris and ELE, which in turn closely match the structural and kinematical properties of the MW disk \citep{Bird:2013aa}. 

Despite the differences between the three simulations, we found that hot gas constitutes nearly 1\% of the total virial mass at $z=0.5$ regardless of the thermodynamical model. The close confinement of the hot gas (within 100~kpc at that redshift) goes hand in hand with the morphology of that phase presented in the temperature maps at $z=0.5$ of Fig.~\ref{fig:fig1}, justifying the usage of the term \emph{corona}.  The hot coronal gas is surrounded by an envelope of warm--hot gas that extends beyond the virial radius of a galaxy.
 

The warm--hot and hot gas together are of the order of stellar mass, and constitute as much as 72--80\% of the total gas content.  As they correspond to the matter in the temperature regime, which to date remains incompletely mapped, this unequivocally points to the possibility of the existence of undetected large reservoir of matter around spiral galaxies.  It therefore favors the already known claim \citep{Nicastro:2005aa} that the 
warm--hot medium is a potential solution to the missing baryon problem. 

COS--Halos results imply that most of the missing baryons actually reside within the virial radius of spiral galaxies {,} which is consistent with our findings. Indeed, all variations of Eris contain 67--74\%  of expected baryon budget at $z=0.5$ (fractions in the same range are found at $z=0$ for Eris and ELE). The relative contributions of the warm--hot and hot medium to the baryon budget are  {also} consistent with, although closer to the upper end of, the range inferred for $L^* $ galaxies by \cite{Werk:2014}. The latter work also reports a large amount of warm and cold gas, larger than both what we find in our simulations and what is found in the Milky Way, even after {  accounting for the mass in the MW halo likely missed by observers. The correction derived by \cite{Zheng:2015} is premised on the fact that they find a good agreement between the kinematics of the CGM gas in their simulations and the observations of the CGM in COS-Halos. Our simulation results are not subject to this correction because it is not necessary for haloes viewed from the outside.}

{  \cite{Zheng:2015} reports that the HI mass of HVCs (high velocity clouds) corrected for the velocity obscuration and the HI mass of the high--velocity gas in the Magellanic System (excluding Magellanic Clouds) scales as $M_{HI} = 3\times 10^8(d[kpc]/55)^2M_{\odot}$ \citep[see also][for the details on the origin of that relation]{Putman:2003}. This in turn implies: $2.5\times 10^{9}$~M$_{\odot}$ at 160~kpc, as opposed to $2\times 10^{10}$~M$_{\odot}$ reported by COS-Halos based on estimates derived from HI surveys and baryonic Tully-Fisher relation \citep[see for the reference:][]{Martin:2010, McGaugh:2012, Werk:2014}.  In our simulations, the mass of HI at 160~kpc amounts to $1.5-3\times10^9$~M$_{\odot}$ and thus is in agreement with the Milky Way HI results. }

Our results also suggest that either our MW-like galaxy formed in a region that was rendered somewhat baryon deficient by winds from smaller systems, from which our final galaxy was assembled \citep[c.f.][]{Lichen:2015, Ford:2013} or that the remaining missing baryons were vented out of the dark halo by powerful supernovae--driven outflows. The fact that the first--generation runs assembled up to 93\% of cosmic mean in baryons within 3 virial radii implies that the missing baryon problem is a scale--dependent problem, which would be solved by extending measurements sufficiently far from the virialized region of a halo.  

Since the warm--hot medium is shown to be sensitive to the details of the simulation setup, measuring its density and entropy profiles in several nearby galaxies would yield important constraints on sub--grid models for galaxy formation, in contrast to that of hot coronal gas. The hot gas seems to be much less sensitive to significant changes in feedback strength in our simulations in terms of its extension and abundance. Once reliable spatial mapping of the warm--hot medium becomes available, in particular from near the galactic disk to beyond the virial radius, the baryon fraction as a function of radius will provide a stringent test for feedback models (Table~\ref{tab:tab2}). 

{  Recently,} \cite{Marinacci:2014G} {  considered 8} gaseous halos of Milky Way-sized objects. Their temperature cuts of gas phases are similar to ours and some of their findings align well with the conclusions of this work. Namely, the cumulative mass profiles of the galaxies studied by \cite{Marinacci:2014G} (see their Fig. 3) show clearly that their hot gas phase is confined within the inner halo, between 50 and 100~kpc, while their warm--hot phase extends beyond the virial radius and in most cases dominates the baryon budget of the diffuse halo. For these two phases, thus, results are in line with ours. However, their galaxies show a {   cool ($<10^5 K$) } phase, which often extends beyond the region of the hot corona and dominates over the latter in the baryon budget. This latter difference may be traced to differences in the feedback model as well as in the numerical hydrodynamics technique (Lagrangian mesh vs. SPH).  

{  For example, on the one hand the {\sc Gasoline} implementation of the SN feedback prevents a part of the heated gas from cooling on a Myr scale \citep{Stinson:2006aa}, which could in principle facilitate retaining higher temperatures by the halo gas while at the same time depleting the reservoir of cool gas. On the other hand, the kinetic feedback implementation of Dav{\'e} and collaborators \citep[e.g.][]{Finlator:2008, Oppenheimer:2008a, Ford:2013, Lichen:2015} and \cite{Vogelsberger:2013}
entails a component of hydrodynamically decoupled winds with cold ejecta. In the {\sc Arepo} simulations \citep{Vogelsberger:2013}, such winds have been found to trigger the formation of cold string--like features in the CGM.  To prevent artefactual star formation in these regions, \cite{Marinacci:2014} split the energy given to the wind into equal thermal and kinetic parts, thus solving the problem with warmer ejecta. It has been shown that the intermediate gas phase between the low-density medium at temperatures around $10^{5.5}$~K and the dense cold gas in the proximity of a galaxy at temperatures of order $10^4$~K is causally connected with the ejecta \citep{Marinacci:2014G}, although AGN feedback and the cooling of the warm--hot gas are also at play. Thus, both the thermalization of the galactic wind energy and the cooling shut--off would have a non--negligible impact on the difference in the total cool ($<10^5$~K) mass budget of \cite{Marinacci:2014G} gaseous galactic halos and ours. We stress that these conjectures put forward here require actual testing by a direct comparison of the two types of simulations.}

In this paper special attention has been given to the major constraints available on the Milky Way gaseous halo. Firstly, we computed X--ray luminosities of coronae in the 0.5--2~keV band.  The X--ray luminosity of E2k is much higher than that of Eris or ELE because the coronal gas of E2k is hotter, denser and more metal enriched, which in turn boosts the radiative cooling rate. Moreover, E2k has a significantly higher star formation rate at $z<1$ compared to Eris and ELE (Mayer et al. in prep.). The latter could be the result of both the stronger feedback delaying SF to low $z$ and the metal cooling allowing for attaining higher star formation rates once the feedback becomes weaker. A very bright X--ray halo is thus likely connected with being still vigorously star forming, a state which is at odds with what is found for the Milky Way. Nevertheless, the X--ray luminosity of E2k lies within the range observed in normal galaxies.

{  Although one simulation, E2k, has an X-ray luminosity that is comparable with that of very massive spirals, its $L_X$ is 2 orders of magnitude higher than what is observed for the Milky Way. However, the other two runs are instead consistent with a number of constraints on the corona of the MW.} In particular, their $L_X$ is only a factor of 2 (Eris) and 5 (ELE) larger than the value for the $Z=0.3$~Z$_{\odot}$ corona reported by \cite{Miller:2014aa}. This difference likely reflects the underlying assumption of the constant metallicity and a cooling flow in the calculation of $L_X$ performed by \cite{Miller:2014aa}. Our results are free of these {  limitations}.

Secondly, we were also able to fit well the spherical $\beta$ model of \cite{Miller:2013aa} to the electron number densities in our simulations, although the gas structure in the region within twice the radius of the disk turns out to be sensitive to the choice of the feedback model. More explicitly, very energetic feedback of E2k gives rise to the discrepancy in {  the electron density profile in} the region up to 30~kpc that we attribute to both the higher ionization fraction and higher metallicity near the disk. 

The agreement of the simulations with refined M$\&$B model implies that electron number densities are rather unaffected by feedback strength beyond {  a radial distance of 30~kpc} in contrast to the central part of the halo. A caveat is that all our runs adopted the same feedback sub--grid model, namely, blastwave feedback. Future investigations with radically different feedback models,
such as the new superbubble feedback developed in GASOLINE2 \citep{Woods:2014}, will be mandatory. 

The growing body of observational work goes now beyond the OVII absorption measurements analyzed in this paper. 
\cite{Miller:2014aa} recently published a similar analysis for OVII emission that is in good agreement with our simulations as well, though outside the galactic disk.  The authors in the latter paper also report the independent measurements of OVIII emission, which are consistent with the OVII absorption model.  We recall that in this paper we focused on the results of \cite{Miller:2013aa} for the optically thin corona, although \cite{Miller:2013aa} inspect the saturated case as well. Note that \cite{Miller:2014aa} report a shallower slope $\beta$ for the OVII emission line measurements than for the absorption line results in the optically thick regime. This discrepancy is expected  due to a combination of their treatment of optical depth corrections and the overall variation in the OVII emission lines. It is because of the latter uncertainties, as well as the requirement of consistency accompanying computation of ionization fractions with CLOUDY, that we 
decided to focus our analysis on their optically thin model.

The total number densities of coronal gas are compatible with the number density inferred from \cite{Gatto:2013aa}, based on the ram--pressure stripping argument for dwarf galaxy satellites. While in their work authors explicitly focus on the {  single-phase}, isothermal hot gas only, the constraints are general because ram pressure depends on the density of gas and the velocity of a dwarf, not the temperature of gas.  We thus conclude that our halos could have stripped Sextans and Carina, as well as most MW dSphs with orbits coming within 100~kpc.  Although the number densities of our simulations are slightly higher than the estimate from the ram pressure stripping of LMC, this discrepancy could stem from the fact that \cite{Salem:2015} use the \cite{Miller:2013aa} best--fit without accounting for the metallicity gradient in gaseous halos. In fact, such correction would make the number density profile shallower, hence closer to our results. Moreover, the tidal interactions between the LMC and SMC must play a dominant role in shaping the gaseous Magellanic System, likely causing some of the gas stripping. Not accounting for this effect places an important limitation on the model of \cite{Salem:2015}. All in all, our results lend support to models, in which the combination of tidal shocks
and ram pressure stripping alone explains the origin of dSphs as the descendants of an early in-falling population of gas--rich dwarfs \citep{Mayer:2007}.

With the increasing adequacy of cosmological, hydrodynamical simulations of MW--like galaxies, we have reached the era when the gaseous galactic halos have become realistic. Thus with this work we fulfilled the key requirement to justify the future usage of the Eris suite of simulations in order to understand in depth the physical nature, assembly and evolution of the coronal gas, which will be the focus of Paper II. 
 As suggested by the present--day entropy distributions, distinct gas phases likely had different thermodynamical histories. 
Therefore extending our work to multiple tracers, especially for the warm-hot and warm phase, such as CII, SiIII, MgII \citep[e.g.][]{Herenz:2013}, as soon as the sensitivity of 
observational instruments enables mapping of their spatial distributions, will be of tremendous value to constrain the physics implemented
in the simulations.

\section{Acknowledgements}
Authors would like to thank Ali Rahmati and Filippo Fraternali for invaluable comments regarding the manuscript, Romeel Dav{\'e} and Lichen Liang for helpful discussions regarding the X-ray luminosity of hot halo gas, and Matthew Miller and Joel Bregman for help in understanding their results. Support for this work was provided by the NSF through grants OIA-1124453 and AST-1229745,  {and by NASA through grants NNX12AF87G  and HST-AR-13904.001-A (P.M.)}  A.B. is supported by NSERC Canada through the Discovery Grant program.

\bibliographystyle{apj}
\bibliography{apj-jour,references}


\end{document}